\documentclass[letterpaper]{jpconf}
\usepackage{epsfig}

\def\Journal#1#2#3#4{#4 {#1}{\bf #2} #3 }

\def\IJMPA{{\it Int. J. Mod. Phys.}~{A}\,}

\def\NPB{{\it Nucl. Phys.}~{B}\,}
\def\PLB{{\it Phys. Lett.}~{B}\,}
\def\PLC{\it Phys. Repts.\ }
\def\PRL{\it Phys. Rev. Lett.\ }

\def\PRD{{\it Phys. Rev.}~{D}\,}

\def\ARNS{{\it Ann. Rev. Nucl. Part. Sci.\ }} 
\def\RMP{\it Rev. Mod. Phys.\ }
\begin{document}
\title[From the ISR to RHIC---2-particle correlations]{From the ISR to RHIC---measurements of hard-scattering and jets using inclusive single particle production and 2-particle correlations}

\author{M J Tannenbaum\footnote[1]{Supported by the U.S. Department of Energy,  Contract No. DE-AC02-98CH1-886 .}}

\address{Physics Dept., 510c, Brookhaven National Laboratory, Upton, NY 11973-5000 USA}
\ead{mjt@bnl.gov}




\begin{abstract}
Hard scattering in p-p collisions, discovered at the CERN 
ISR in 1972 by the method of leading particles, proved that the partons of Deeply Inelastic Scattering strongly interacted with each other. Further ISR measurements utilizing inclusive 
single or pairs of hadrons established that high 
$p_T$ particles are produced from states with two roughly back-to-back jets which are 
the result of scattering of constituents of 
the nucleons as described by Quantum Chromodynamics (QCD), which was developed during the course of these measurements.  
These techniques, which are the only practical method to study hard-scattering and jet 
phenomena in Au+Au central collisions at RHIC energies, are reviewed, as an introduction to present RHIC measurements. 
\end{abstract}

\section{Introduction}

  In 1998, at the QCD workshop in Paris, Rolf Baier asked me whether jets could be measured in Au+Au collisions because he had a prediction of a QCD medium-effect on color-charged partons traversing a hot-dense-medium composed of screened color-charges~\cite{BaierQCD98}. I told him~\cite{MJTQCD98} that there was a general consensus~\cite{Strasbourg} that for Au+Au central collisions at $\sqrt{s_{NN}}=200$ GeV, leading particles are the only way to find jets, because in one unit of the nominal jet-finding cone,  $\Delta r=\sqrt{(\Delta\eta)^2 + (\Delta\phi)^2}$, there is an estimated $\pi\times{1\over {2\pi}} {dE_T\over{d\eta}}\sim 375$ GeV of energy !(!) The good news was that hard-scattering in p-p collisions was originally observed by the method of leading particles. 

   The other good news was that the PHENIX detector had been designed to make such measurements and could identify and separate direct single $\gamma$ and $\pi^0$ out to $p_T\geq 30$ GeV/c~\cite{Shura2000}. It is ironic that the identification of $\pi^0$ and the separation from direct single $\gamma$ out to such a large $p_T$ in PHENIX was primarily driven by the desire to measure the polarized gluon structure function in p-p collisions in the range $0.10\leq x_T\leq 0.30$ via the longitudinal two-spin asymmetry of direct photon production~\cite{MJTQCD98}. This  illustrates that a good probe of QCD in a fundamental system such as p-p collisions, provides a well calibrated probe of QCD in more complicated collisions such as Au+Au, should the need arise, which it fortunately did. In the 1980's when RHIC was proposed, hard processes were not expected to play a major role. However, in 1998~\cite{MJTQCD98}, inspired by Rolf and collaborators, and before them by the work of Gyulassy~\cite{MGyulassy} and Wang~\cite{XNWang}, I indicated that my best bet on discovering the QGP was to 
utilize semi-inclusive $\pi^0$ or $\pi^{\pm}$ production in search for ``high $p_T$ suppression".

   It is not generally realized that hard-scattering was discovered---in both Deeply Inelastic lepton scattering (DIS) at SLAC~\cite{DIS} and in high $p_T$ particle production in p-p collisions~\cite{CCR,SS,BS} at the CERN ISR---before the discovery of QCD. During the period 1972-1982, it was proved by single inclusive and two-particle correlation measurements that high $p_T$ particles are produced in p-p collisions from states with two roughly back-to-back jets which are the result of scattering of constituents of the nucleons, at a rate much higher than was predicted from coulomb scattering~\cite{BBK}. {\em This indicated that the partons of DIS interacted strongly ($\gg EM$) with each other.}  The measurements were consistent in detail with QCD, which was developed during this period. In that era, scaling laws relating the transverse momentum ($p_T$) spectra at different values of center-of-mass (c.m.) energy, $\sqrt{s}$, were the key to understanding the underlying physics. Absolute cross sections played a minimal role. Measurements of high $p_T$ particle production near mid-rapidity, at symmetric p-p or A+A colliders, are ideal for such scaling studies---most systematic errors from varying the $\sqrt{s}$ or the species cancel since the detectors are at rest near $90^{\circ}$ in the c.m. system of the reaction and are thus insensitive to longitudinal effects. 
   
\section{Systematics of single particle inclusive production in p-p collisions.}
\label{sec:single}
\begin{figure}[b]
\begin{minipage}{18pc}
\includegraphics[width=18pc]{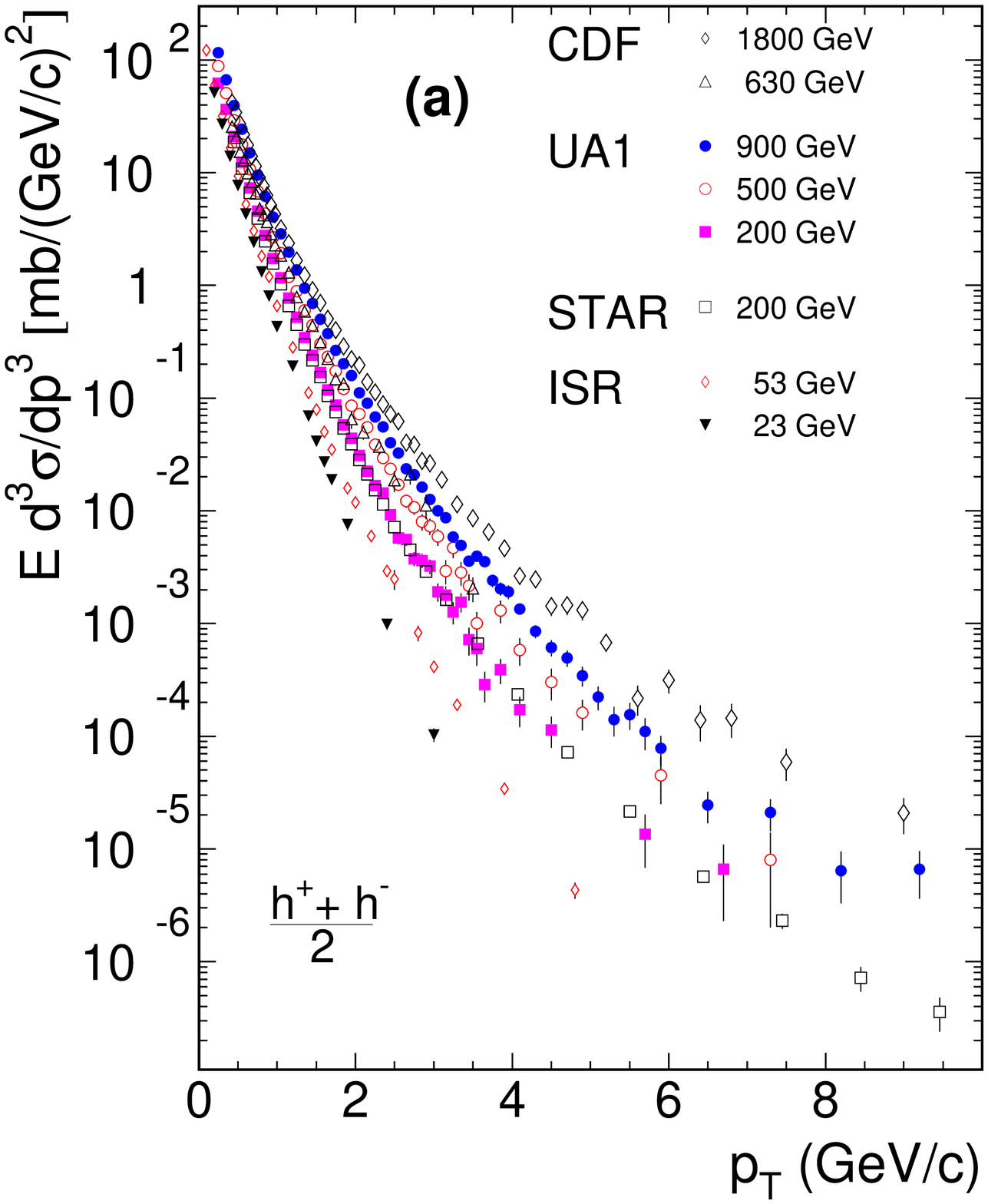}
\caption[]{$E {d^3\sigma}/{dp^3}$ vs. $p_T$ at mid-rapidity as a function of $\sqrt{s}$ in $p-p$  and $p-\overline{p}$ collisions. }
\label{fig:mjt-hpT}
\end{minipage}\hspace{2pc}%
\begin{minipage}{18pc}
\includegraphics[width=18pc]{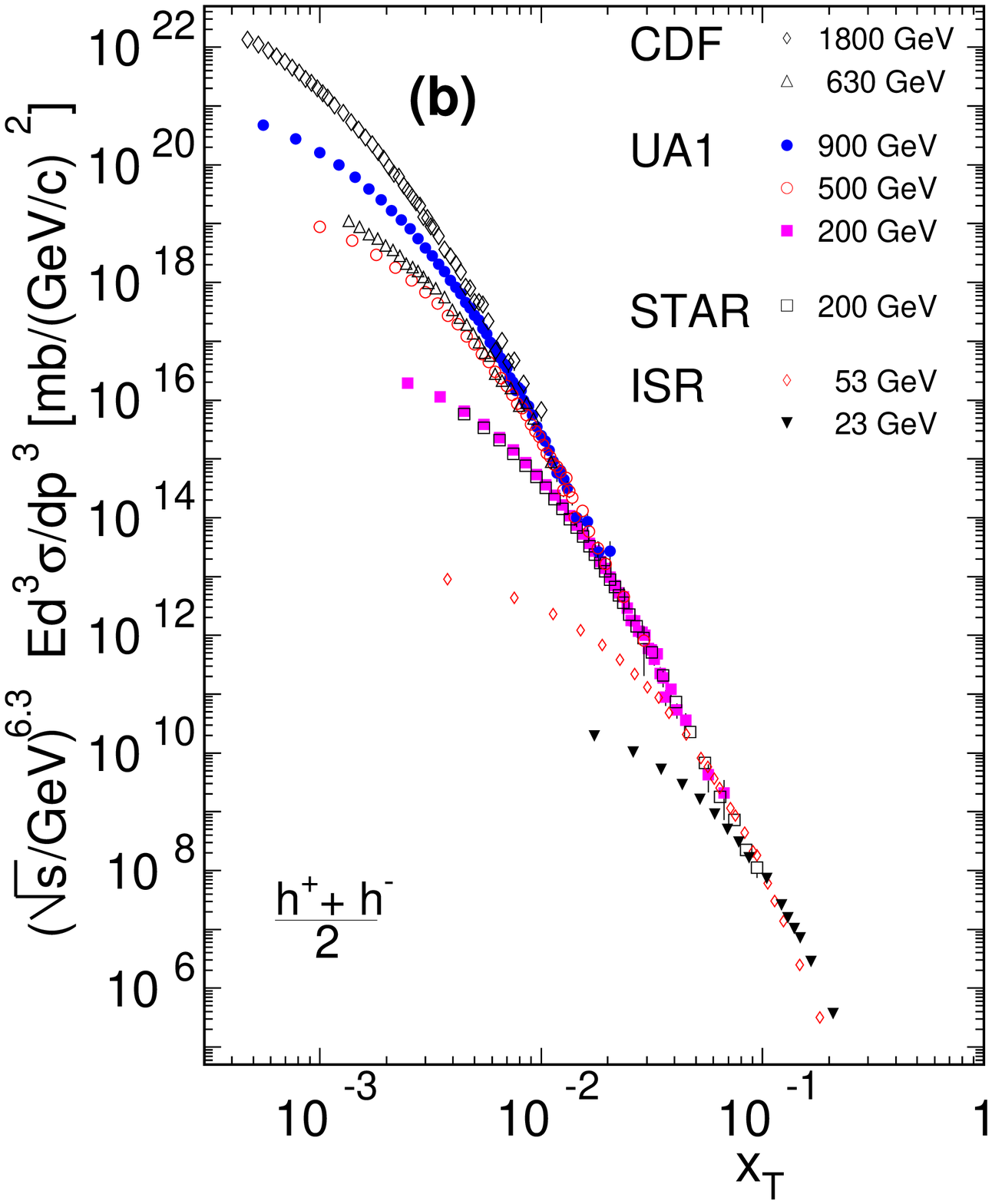}
\caption[]{Data from Fig.~\ref{fig:mjt-hpT} plotted as $\sqrt{s}({\rm GeV})^{6.3}\times Ed^3\sigma/dp^3$ vs. $x_T=2{p_T}/\sqrt{s}$.  }
\label{fig:mjt-hxT}
\end{minipage} 
\end{figure}

In p-p collisions, the invariant cross section for non identified 
charge-averaged hadron production at 90$^\circ$ in the c.m. system as a 
function of the transverse 
momentum $p_T$ and c.m. energy $\sqrt{s}$ has a characteristic 
shape (Fig.~\ref{fig:mjt-hpT}). There is an 
exponential ($e^{-6p_T}$) at low $p_T$, which depends very little on 
$\sqrt{s}$. This is the soft physics region, 
where the hadrons are fragments of 
`beam jets'. At higher $p_T$, there is a power-law tail which depends very 
strongly on $\sqrt{s}$. 
This is the hard-scattering region, where the hadrons 
are fragments of the high $p_T$ QCD jets from constituent-scattering. 

	The hard scattering behavior for the reaction $p+p\rightarrow C+X$ is easy to understand from general principles proposed by Bjorken and collaborators~\cite{Bj,BBK} and subsequent authors~\cite{CIM,CGKS}.  Using the principle of factorization of the reaction into parton distribution functions for the protons, fragmentation functions to particle $C$ for the scattered partons and a short-distance parton-parton hard scattering cross section, the invariant cross section for the inclusive reaction, where particle $C$ has transverse momentum $p_T$ near mid-rapidity, is given by the general `$x_T$-scaling' form~\cite{CIM}, where $x_T=2p_T/\sqrt{s}$: 
\begin{equation}
E \frac{d^3\sigma}{dp^3}=\frac{1}{p_T^{n}} F({2 p_T \over \sqrt{s}})
= \frac{1}{\sqrt{s}^{\,n}} G({x_T}) .
\label{eq:mjt-bbg}
\end{equation}
The cross section has 2 factors, a function $F$ $(G)$ which `scales', i.e. depends only on the ratio of momenta; and a dimensioned factor, ${p_T^{-n}}$   $(\sqrt{s}^{\,-n})$,   
where $n$ depends on the quantum exchanged  
between constituents. For QED or vector-gluon exchange~\cite{BBK}, $n=4$, analogous to the $1/q^4$ form of Rutherford Scattering, while  $n$=8 for the the constituent interchange model (CIM)~\cite{CIM}, quark-meson scattering by the exchange of a quark. When QCD is added to the mix~\cite{CGKS}, pure $n=4$ `vector-gluon' scaling breaks down and $n$ varies according to the $x_T$ and $\sqrt{s}$ regions used in the comparison, $n\rightarrow n(x_T, \sqrt{s})\sim 4-6$.  

	It is evident from Fig.~\ref{fig:mjt-hpT} that hard-scattering, which is a relatively small component of the $p_T$ spectrum at $\sqrt{s}\sim 20$ GeV, dominates for $p_T\geq 2$ GeV/c by nearly 2 orders of magnitude at RHIC c.m. energies compared to the soft physics $e^{-6p_T}$  extrapolation. The characteristic $\sqrt{s}$ dependence of the high $p_T$ tail is simply explained by the $x_T$ scaling of the spectrum.  
However, it is worthwhile to note that it took quite some time for $x_T$ scaling with the value of $n=5.1\pm 0.4$, consistent with QCD, to be observed at the CERN-ISR~\cite{CCR,CCRS,CCOR}. This was due to the so-called `intrinsic' transverse momentum of partons, the ``$k_T$-effect'', which causes a transverse momentum imbalance of the outgoing parton-pairs from hard-scattering, making the jets not exactly back-to-back in azimuth. The ``$k_T$-effect'' acts to broaden the $p_T$ spectrum, thus spoiling the $x_T$-scaling at values of $p_T\leq 7.5$ GeV/c, at the ISR, and totally confusing the issue at fixed target incident energies of 200--400 GeV ($\sqrt{s}\sim 20$ GeV)~\cite{Cronin} due to the the relatively steep $p_T$ spectrum (see Fig.~\ref{fig:mjt-hpT}), which results in a relatively strong broadening effect.     
\section{Theory and Experiment, c. 1972-82} 
\begin{figure}[ht]
\begin{center}
\begin{tabular}{cc}
\psfig{file=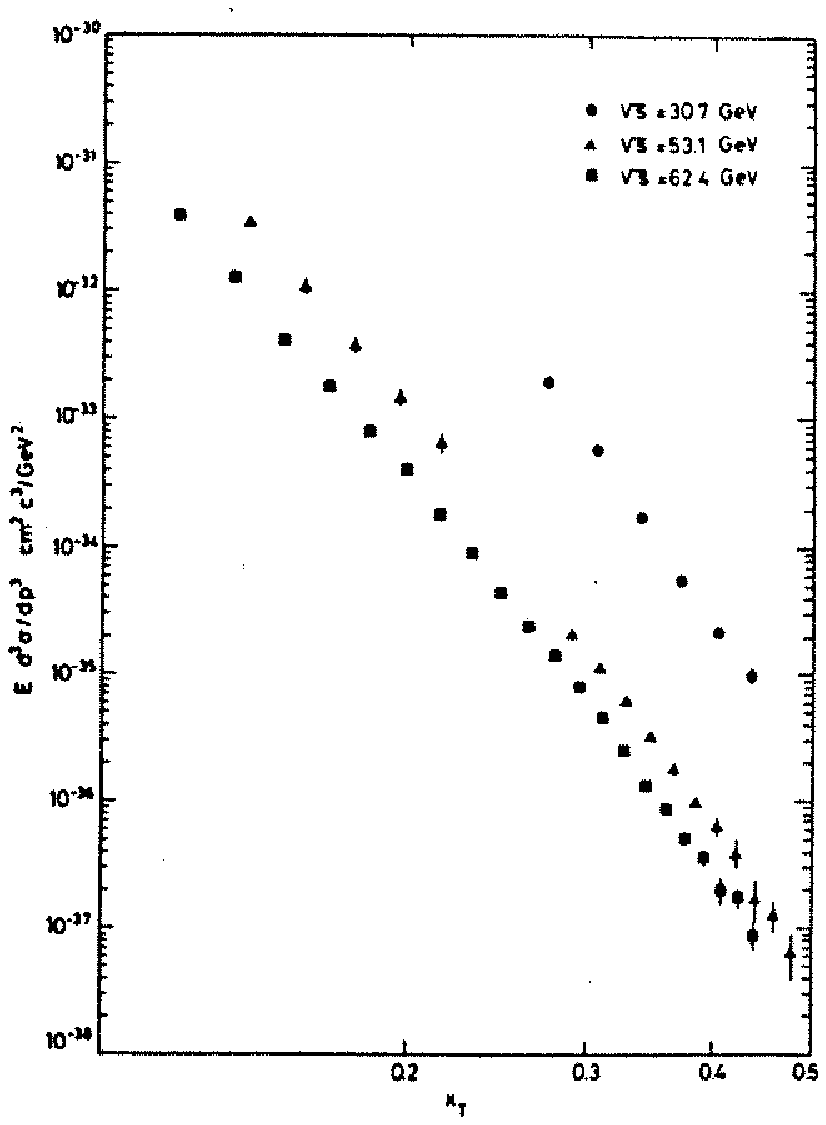,height=3.85in}
\hspace*{0.55in}
\psfig{file=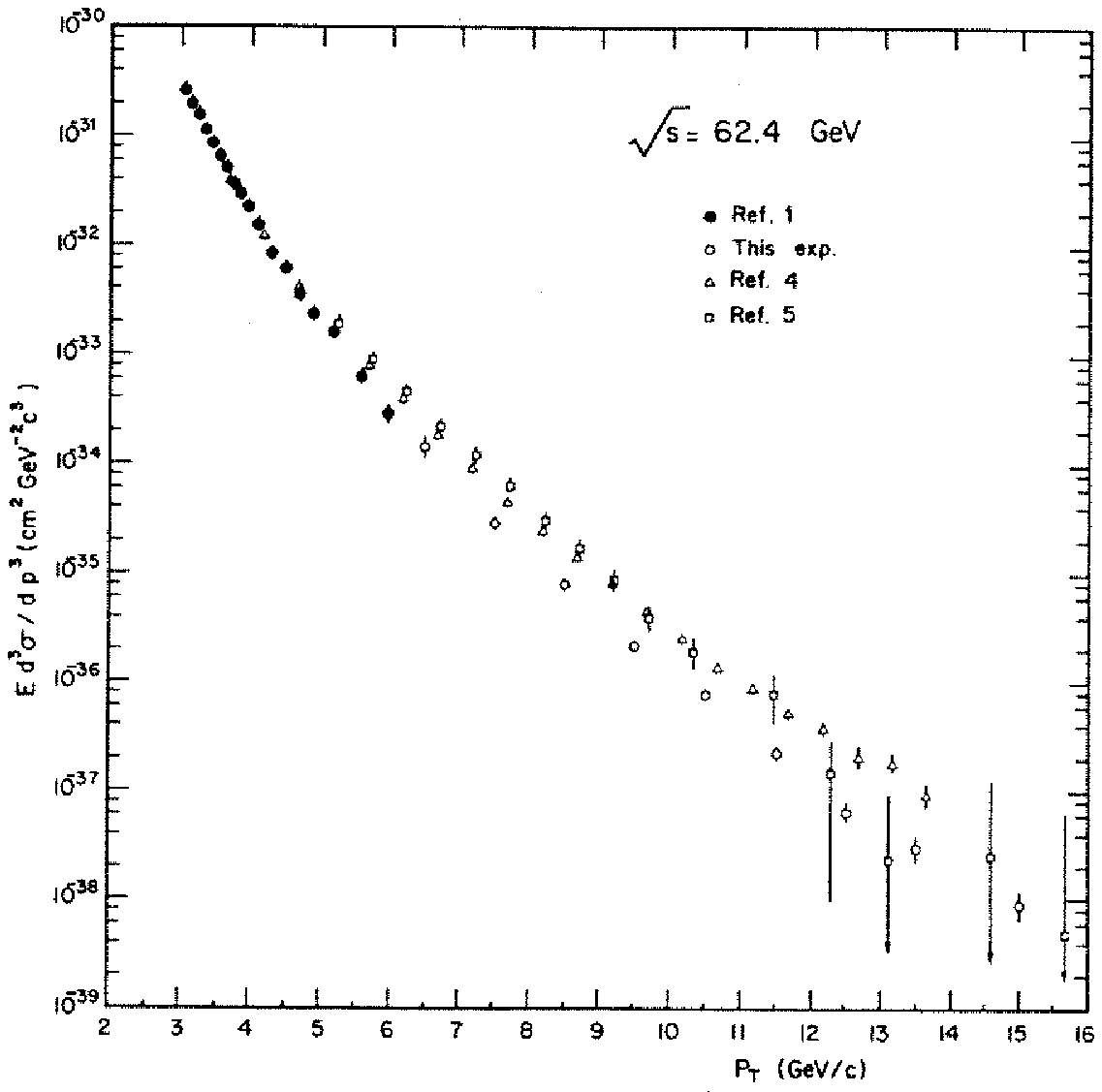,height=3.85in,width=2.75in}\\
\hspace*{0.25in}
\psfig{file=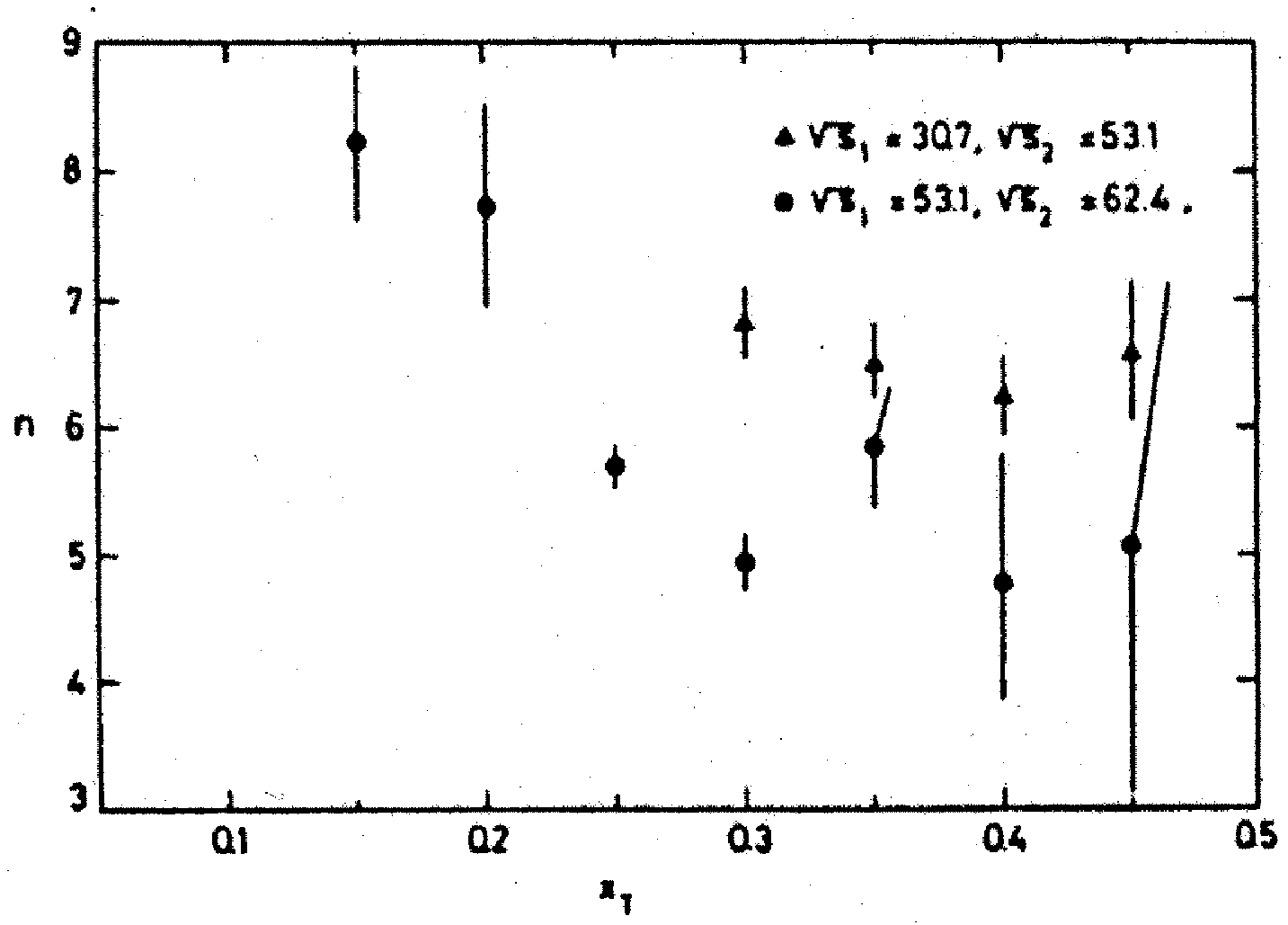,height=1.65in}\hspace*{0.80in}
\psfig{file=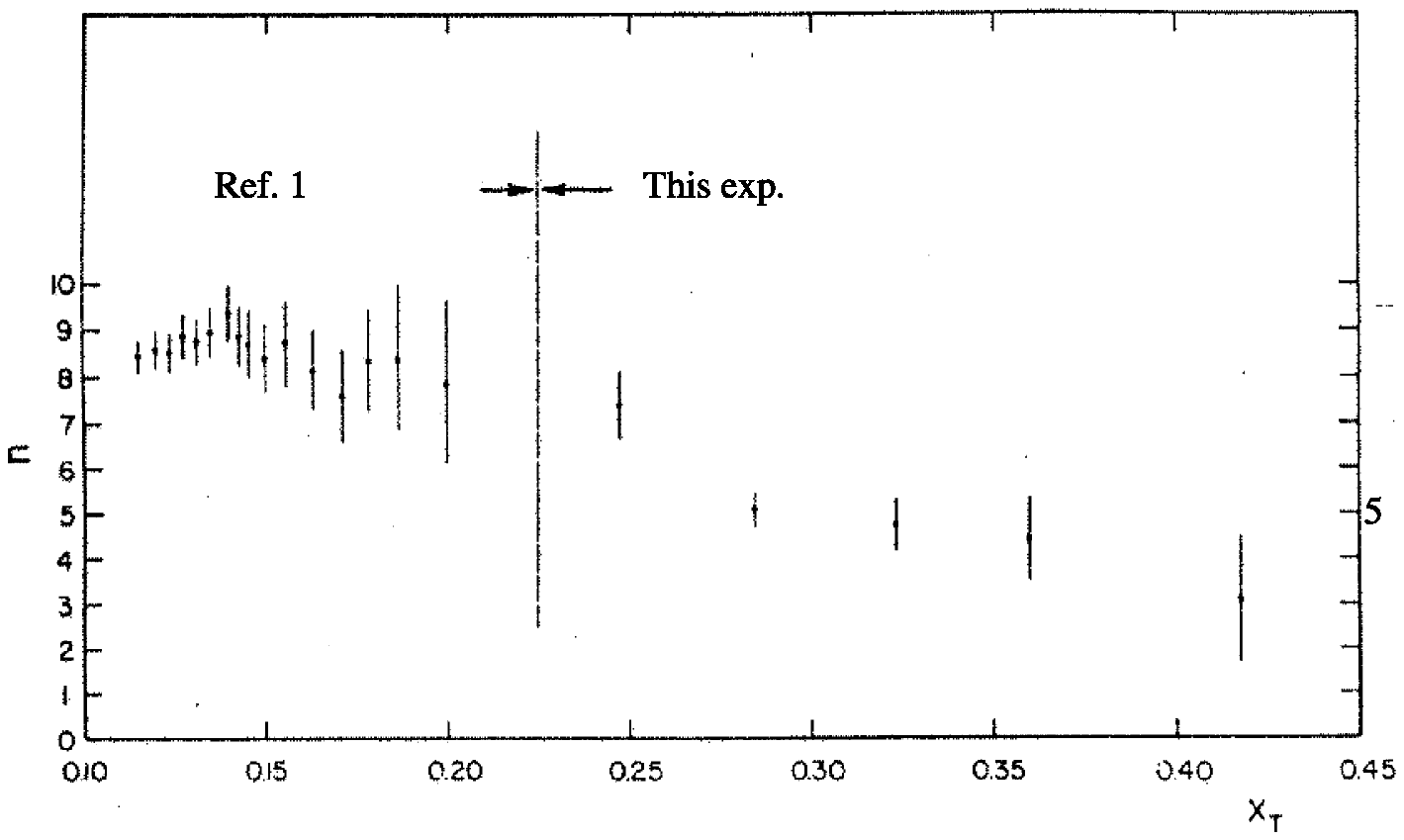,height=1.65in}
\end{tabular}
\end{center}
\vspace*{-1.5mm}
\caption[]
{a) (top-left) Log-log plot of CCOR invariant cross section vs $x_T=2 p_T/\sqrt{s}$; 
b)~(bottom-left) $n(x_T,\sqrt{s})$ derived from the combinations indicated.  
There is an additional common systematic error of 
$\pm 0.33$ in $n$. 
c) (top-right)  Invariant cross section for $\pi^0$ inclusive for several ISR 
experiments, compiled by ABCS Collaboration. d) (bottom-right) $n(x_T,\sqrt{s})$ 
from ABCS 52.7, 62.4 data only. There is an 
additional common systematic error of $\pm0.7$ in $n$.
\label{fig:mjt-ccorxt} }
\end{figure}

    The distortion of $x_T$ scaling from the QCD prediction due to the $k_T$-effect is illustrated by measurements from the CERN-ISR.   
The CCOR measurement~\cite{CCOR} (Fig.~\ref{fig:mjt-ccorxt}a),  with a larger apparatus and much increased integrated 
luminosity, extended the $\pi^0$ measurements of the original high $p_T$ discovery~\cite{CCR,SS,BS} to 
much higher $p_T$. 
In Fig.~\ref{fig:mjt-ccorxt}a, the CCOR $\pi^0$ data for 3 values of $\sqrt{s}$ are plotted vs $x_T$ on a log-log scale. $n(x_T, \sqrt{s})$ is determined for any 2 values of $\sqrt{s}$ by taking 
the ratio of invariant cross sections at fixed $x_T$, with results shown in 
Fig.~\ref{fig:mjt-ccorxt}b: $n(x_T, \sqrt{s})$  clearly varies 
with both $\sqrt{s}$ and $x_T$, it is not a constant. For 
$\sqrt{s}=53.1$ and 62.4 GeV, $n(x_T, \sqrt{s})$ varies from $\sim 8$ at low 
$x_T$ to $\sim 5$ at high $x_T$. The fit~\cite{CCOR}, for $7.5\leq p_T\leq 14.0$ GeV/c,  $53.1\leq \sqrt{s}\leq 62.4$ GeV, is  
\mbox{$E d^3 \sigma/dp^3\simeq p_T^{-{5.1\pm 0.4}} (1-x_T)^{12.1\pm 0.6}$} (including {\em all} systematic errors). 
It is interesting to note that the original ISR high $p_T$ discovery in the $x_T$, $\sqrt{s}$ range where $n=8$, instead of 4, stimulated the development of the constituent interchange model~\cite{CIM}, which seemed to naturally explain the observed value. The best data at FNAL in 1977~\cite{Cronin} also beautifully showed the CIM scaling with $n\simeq 8$ over the range $0.2\leq x_T\leq 0.6$, for 200, 300 and 400 GeV incident energies. However, this effect turned out not to be due to CIM, but to  the broadening by the $k_T$-effect.

	An important feature of the scaling analysis (Eq.~\ref{eq:mjt-bbg}) in  
determining $n(x_T, \sqrt{s})$---{\em is that the absolute $p_T$ 
scale uncertainty and many efficiency and acceptance errors cancel!} The effect of the absoulte scale uncertainty, which 
is the main systematic error in these experiments,  can be gauged from 
Fig.~\ref{fig:mjt-ccorxt}c~\cite{ABCS} which shows the $\pi^0$ cross 
sections from several experiments. The absolute cross sections disagree by 
factors of $\sim 3$ for different experiments but the values of 
$n(x_T, \sqrt{s})$ for the CCOR~\cite{CCOR} 
(Fig.~\ref{fig:mjt-ccorxt}b) and ABCS~\cite{ABCS} experiment 
(Fig.~\ref{fig:mjt-ccorxt}d) are in excellent agreement due 
to the cancellation of the systematic errors in each experiment. Thus, while the individual ISR experiments each provide a data set with common systematic uncertainties which cancel in scaling studies, there is no unique absolute cross section $\pi^0$ measurement from the ISR at 62.4 GeV which can be used as a comparison spectrum for RHIC measurements.

\section{Status of theory and experiment, circa 1982}
 
Hard-scattering was visible both at ISR and FNAL (Fixed Target) energies 
via inclusive single particle production at large $p_T\geq$ 2-3 
GeV/c. Scaling and dimensional arguments for plotting 
data revealed the systematics and underlying physics. The theorists had the 
basic underlying physics correct; but many (inconvenient) details remained to 
be worked out, several by experiment. The transverse momentum 
imbalance of outgoing parton-pairs, the ``$k_T$-effect", was 
discovered by experiment~\cite{CCHK,MJT79}, and clarified by Feynman and collaborators~\cite{FFF}. The first modern QCD calculation and 
prediction for high $p_T$ single particle inclusive cross sections, including 
non-scaling and initial state radiation was done in 1978, by Jeff 
Owens and collaborators~\cite{Owens78} under the assumption that high $p_T$ particles  
are produced from states with two roughly back-to-back jets
which are the result of scattering of constituents of the nucleons (partons). 
   The overall $p+p$ hard-scattering cross section in ``leading logarithm" pQCD   
is the sum over parton reactions $a+b\rightarrow c +d$ 
(e.g. $g+q\rightarrow g+q$) at parton-parton center-of-mass (c.m.) energy $\sqrt{\hat{s}}=\sqrt{x_1 x_2 s}$.  
\begin{equation}
\frac{d^3\sigma}{dx_1 dx_2 d\cos\theta^*}=
\frac{1}{s}\sum_{ab} f_a(x_1) f_b(x_2) 
\frac{\pi\alpha_s^2(Q^2)}{2x_1 x_2} \Sigma^{ab}(\cos\theta^*)
\label{eq:mjt-QCDabscat}
\end{equation} 
where $f_a(x_1)$, $f_b(x_2)$, are parton distribution functions, 
the differential probabilities for partons
$a$ and $b$ to carry momentum fractions $x_1$ and $x_2$ of their respective 
protons (e.g. $u(x_2)$), and where $\theta^*$ is the scattering angle in the parton-parton c.m. system. The characteristic subprocess angular distributions,
{\bf $\Sigma^{ab}(\cos\theta^*)$},
and the coupling constant,
$\alpha_s(Q^2)=\frac{12\pi}{25} \ln(Q^2/\Lambda^2)$,
are fundamental predictions of QCD~\cite{CutlerSivers,Combridge:1977dm}.

\begin{figure}[ht]
\begin{center}
\begin{tabular}{cc}
\hspace*{-0.1in}\includegraphics[width=0.75\linewidth]{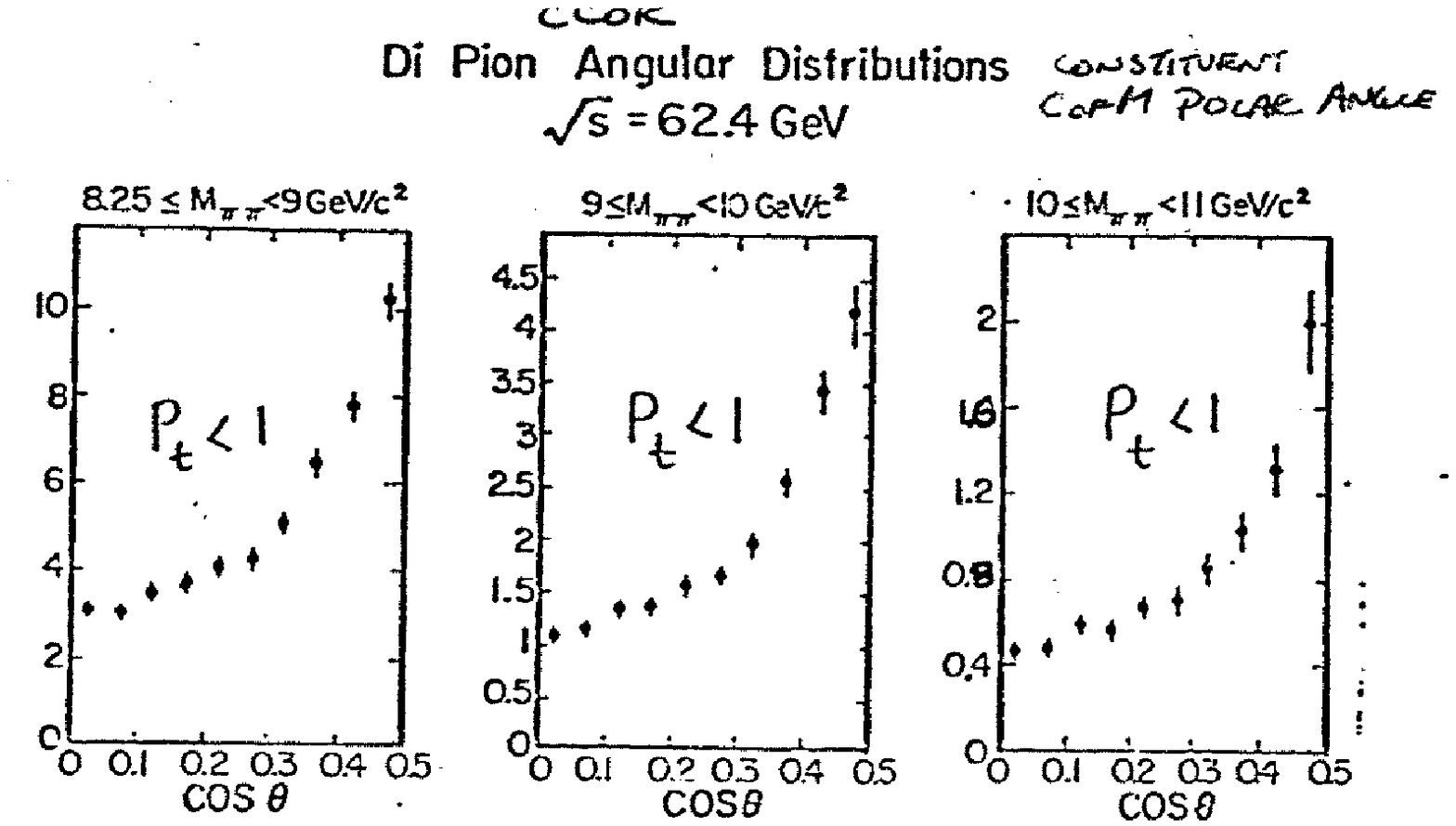} &
\hspace*{-0.35in}\includegraphics[width=0.288\linewidth]{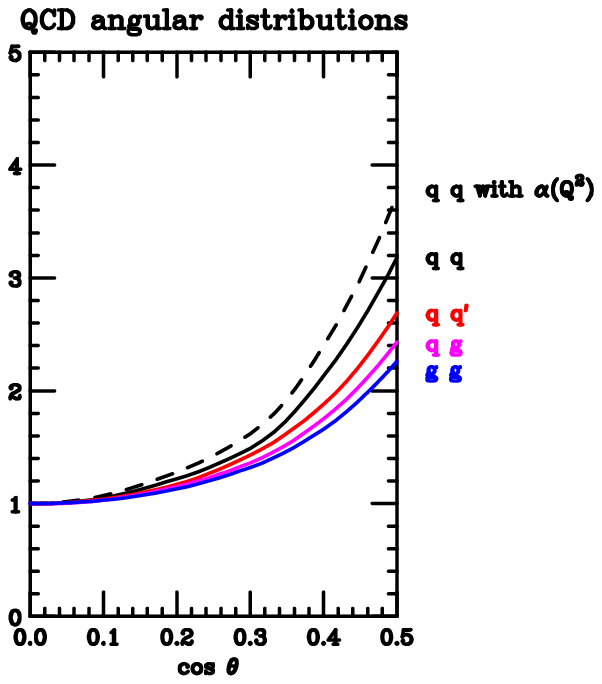}
\end{tabular}
\end{center}
\caption[]
{a) (left 3 panels) CCOR measurement~\cite{Paris82,CCOR82NPB} of polar angular distributions of $\pi^0$ pairs with net $p_T < 1$ GeV/c at mid-rapidity in p-p collisions with $\sqrt{s}=62.4$ GeV for 3 different values of $\pi\pi$ invariant mass $M_{\pi \pi}$. b) (rightmost panel) QCD predictions for $\Sigma^{ab}(\cos\theta^*)$ for the elastic scattering of $gg$, $qg$, $qq'$, $qq$, and $qq$ with $\alpha_s(Q^2)$ evolution.    
\label{fig:mjt-ccorqq} }
\end{figure}

 However, jets in $4\pi$ calorimeters at ISR 
energies or lower are invisible below $\sqrt{\hat{s}}\sim E_T \leq 25$ 
GeV~\cite{Gordon}. Nevertheless, there were many false claims of jet observation in the period 1977-1982 which led to skepticism 
about jets in hadron collisions, particularly in the USA~\cite{MJTIJMPA}. 
A `phase change' in belief-in-jets was produced by one UA2 event 
at the 1982 ICHEP in Paris~\cite{Paris82}, which, together with the first direct measurement of the QCD constituent-scattering angular distribution, $\Sigma^{ab}(\cos\theta^*)$ (Eq.~\ref{eq:mjt-QCDabscat}), using two-particle correlations~\cite{CCOR82NPB}, presented at the same meeting (Fig.~\ref{fig:mjt-ccorqq}), gave universal credibility to the pQCD description of high $p_T$ hadron physics~\cite{Owens,Darriulat,DiLella}.    

\section{Almost everything you want to know about jets can be found using 2-particle correlations.} 

   The outgoing jet-pairs of hard-scattering obey the kinematics of elastic-scattering (of partons) in a parton-parton c.m. frame which is longitudinally moving with rapidity $y=1/2 \ln(x_1/x_2)$ in the p-p c.m. frame. Hence, the jet pair formed from the scattered partons should be co-planar with the beam axis, with equal and opposite transverse momenta, and thus be back-to-back in azimuthal projection. It is not necessary to fully reconstruct the jets in order to measure their properties. In many cases two-particle correlations are sufficient to measure the desired properties, and in some cases, such as the measurement of the net transverse momentum of a jet-pair, may be superior, since the issue of the systematic error caused by missing some of the particles in the jet is not-relevant. A helpful property in this regard is the ``leading-particle effect''. Due to the steeply falling power-law transverse momentum spectrum of the scattered partons, the inclusive single particle (e.g. $\pi$) spectrum from jet fragmentation is dominated by fragments with large $z$, where $z=p_{T\pi}/p_{T_q}$ is the fragmentation variable. The probability for a fragment pion, with momentum fraction $z$, from a parton with $p_{T_q}=p_{T{\rm jet}}$ is:
   \begin{equation}
   {{d^2\sigma_{\pi} (p_{T_q},z) }\over {dp_{T_q} dz }}={{d\sigma_q}\over {dp_{T_q}}}\times D^q_{\pi} (z)={A \over {p_{T_q}^{m-1}}}  \times D^q_{\pi} (z) 
 \qquad ,  \label{eq:mjt-zgivenq}
   \end{equation}
   where $D^q_{\pi} (z)\sim e^{-6z}$ is the fragmentation function. 
The change of variables, $p_{T_q}=p_{T_{\pi}}/z$, ${dp_{T_q}}/{dp_{T_{\pi}}}|_{z}=1/z$, then gives the joint probability of a fragment $\pi$, with  transverse momentum $p_{T_{\pi}}$ and fragmentation fraction $z$: 
\begin{equation}
{{d^2\sigma_{\pi} (p_{T_{\pi}},z)} \over { dp_{T_{\pi}} dz}} 
={A \over {p_{T_{\pi}}^{m-1}}} \times z^{m-2} D^q_{\pi} (z) \qquad . 
\label{eq:mjt-zgivenpi}
\end{equation}
Thus, the effective fragmentation function, given that a fragment (with $p_{T_{\pi}}$) is detected, is weighted upward in $z$ by a factor $z^{m-2}$, where $m$ is the simple power fall-off of the jet invariant cross section (i.e. not the $n(x_T, \sqrt{s})$ of Eq.~\ref{eq:mjt-bbg}~\cite{confusion}). As this property, although general, is most useful in studying `unbiased' away jets using biased trigger jets selected by single particle triggers, it was given the unfortunate name `trigger-bias'~\cite{JacobLandshoff}. 

\begin{figure}[ht]
\begin{center}
\includegraphics[width=0.50\linewidth]{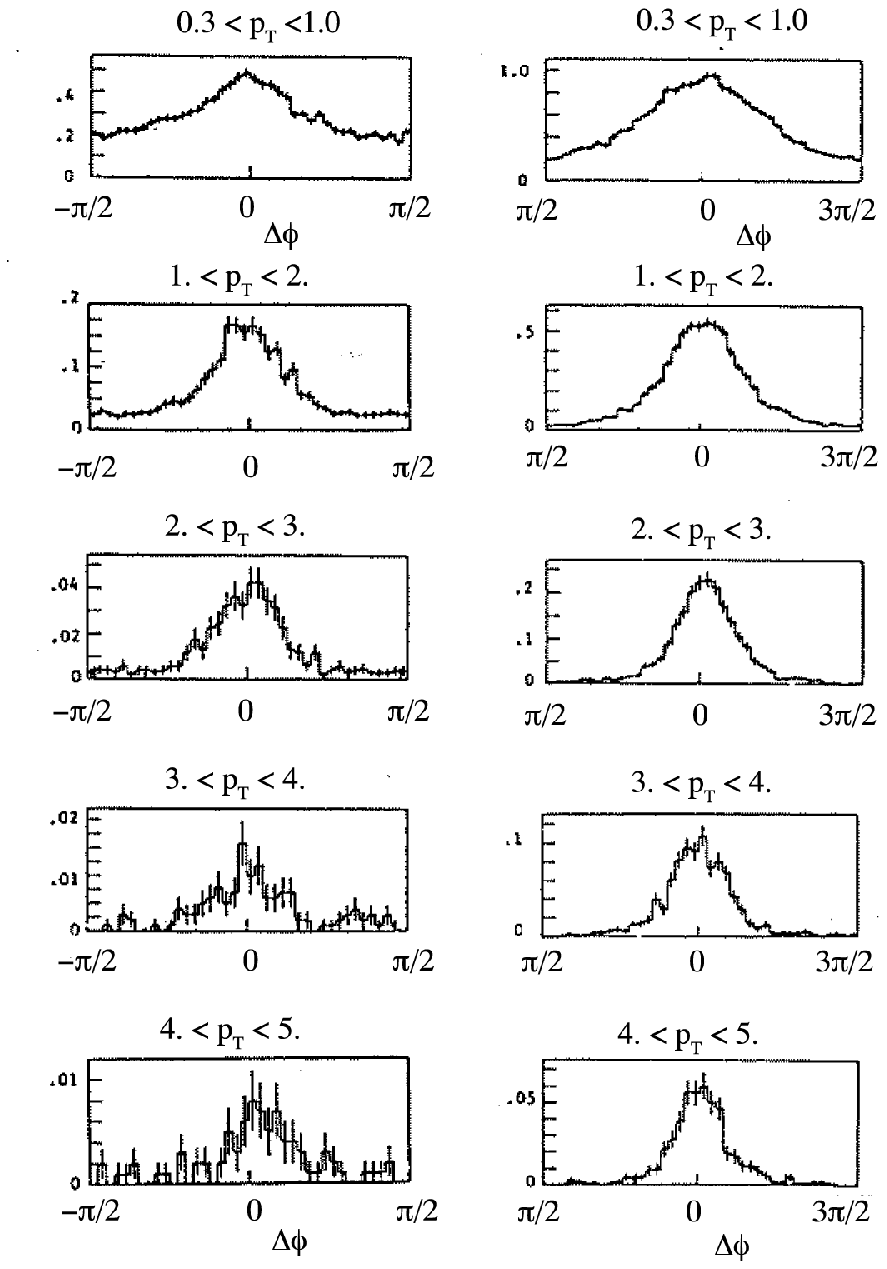} 
\includegraphics[width=0.48\linewidth]{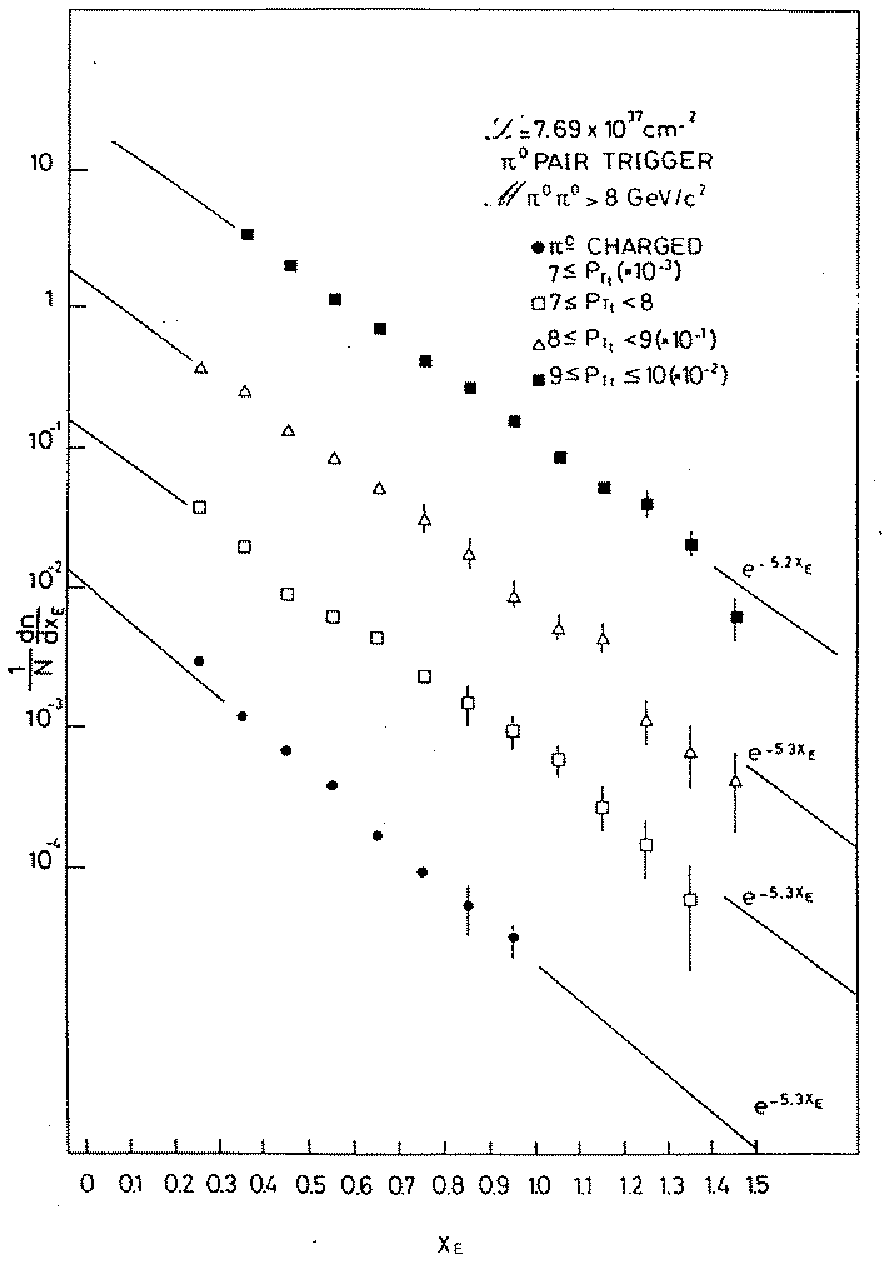}
\end{center}
\vspace*{-0.12in}
\caption[]
{a,b) Azimuthal distributions of charged particles of transverse momentum $p_T$, with respect to a trigger $\pi^0$ with $p_{Tt}\geq 7$ GeV/c, for 5 intervals of $p_T$: a) (left-most panel) for $\Delta\phi=\pm \pi/2$ rad about the trigger particle, and b) (middle panel) for $\Delta\phi=\pm \pi/2$ about $\pi$ radians (i.e. directly opposite in azimuth) to the trigger. The trigger particle is restricted to $|\eta|<0.4$, while the associated charged particles are in the range $|\eta|\leq 0.7$. c) (right panel) $x_E$ distributions (see text) corresponding to the data of the center panel.   
\label{fig:mjt-ccorazi} }
\end{figure}

\begin{figure}[ht]
\begin{center}
\begin{tabular}{cc}
\includegraphics[width=0.47\linewidth]{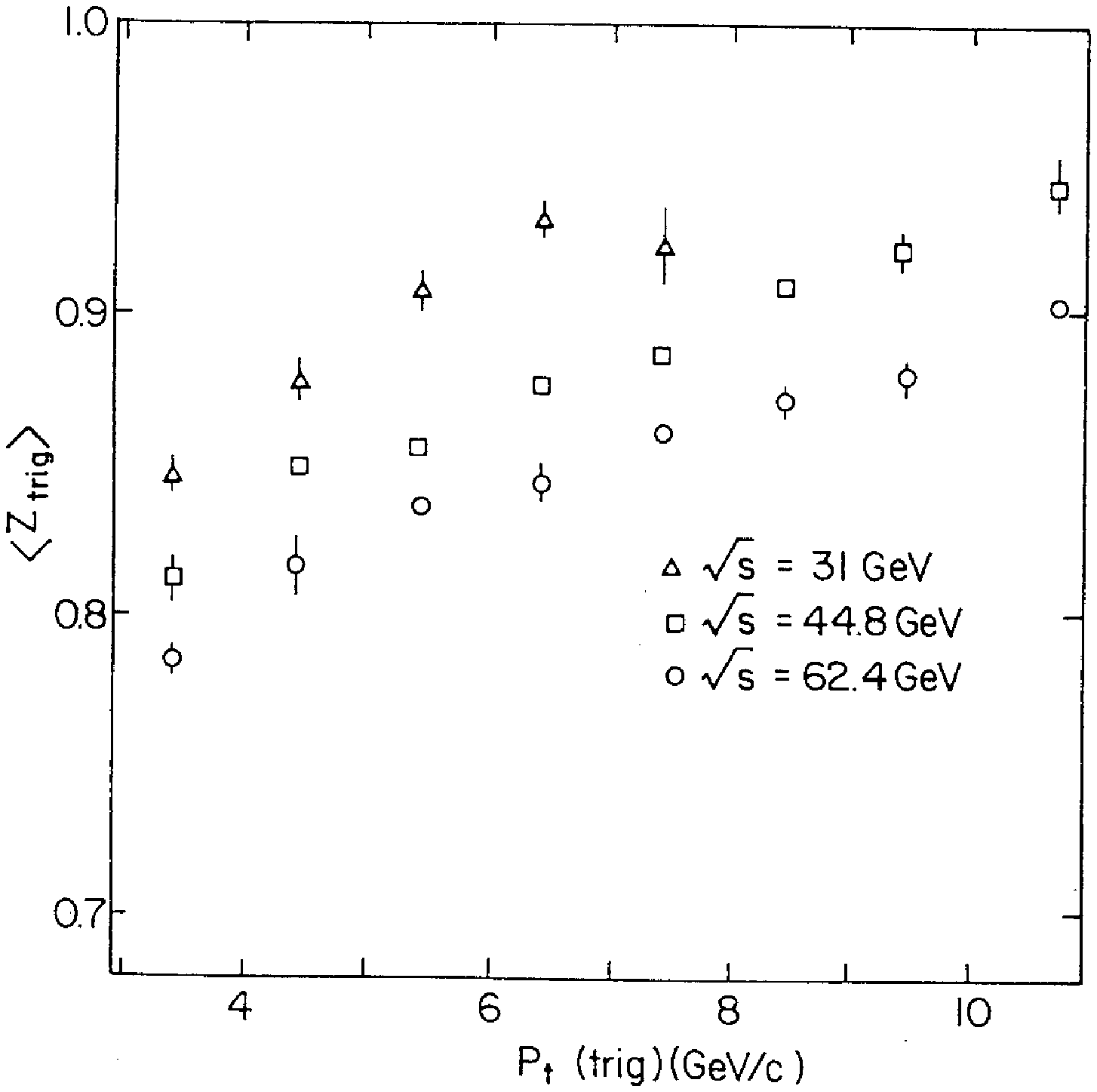} &
\includegraphics[width=0.50\linewidth,angle=-1]{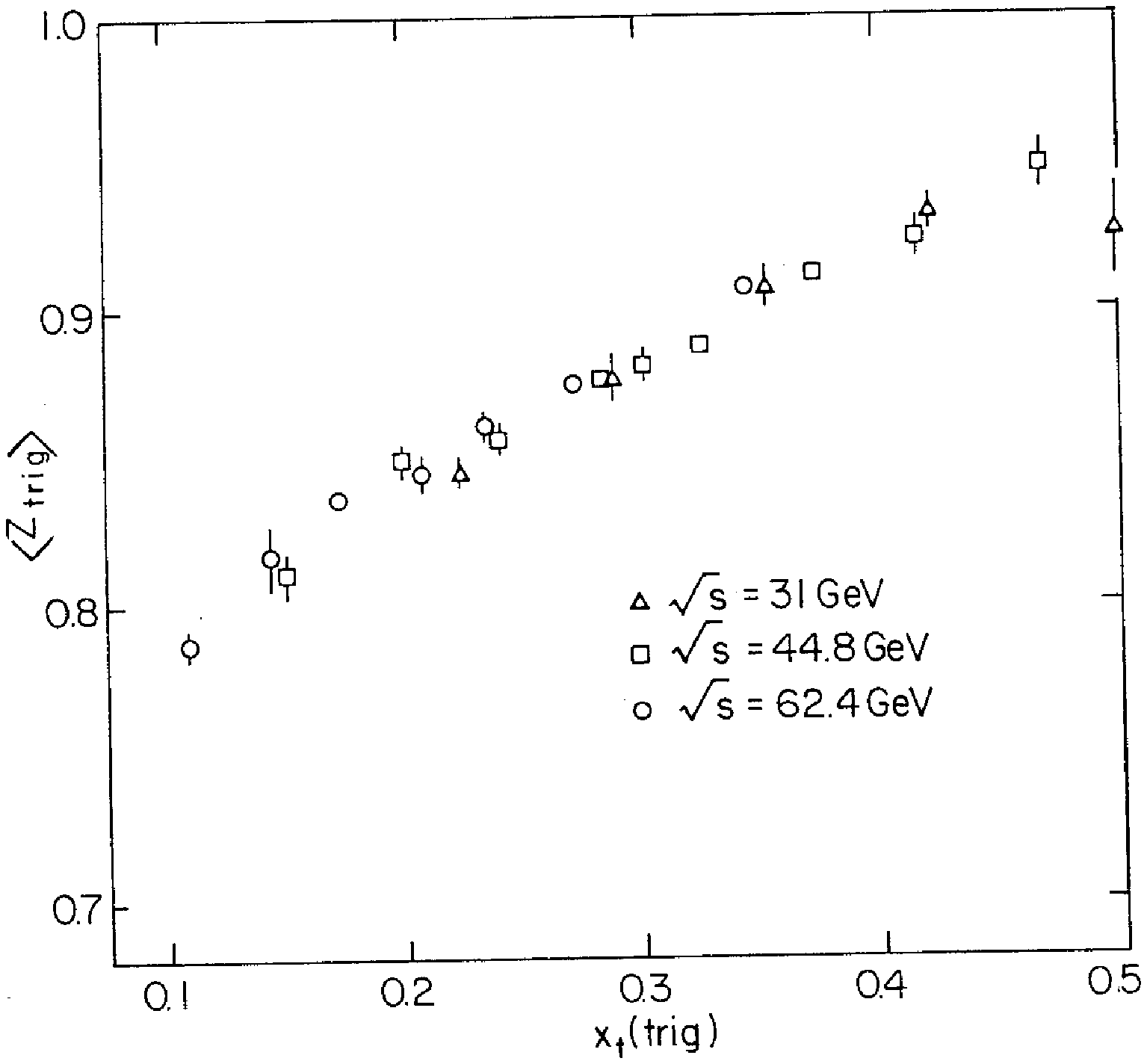}
\end{tabular}
\end{center}
\vspace*{-7mm}
\caption[]
{CCOR~\cite{CCOR82NPB} measurement of $\langle z_{\rm trig}\rangle$ as a function of $p_{Tt}$ (left) and $x_{Tt}=2p_{Tt}/\sqrt{s}$ (right).  
\label{fig:mjt-ccormeanz} }
\end{figure}

   Many ISR experiments provided excellent 2-particle correlation measurements~\cite{Moriond79}. However, the CCOR experiment~\cite{Angelis79} was the first to provide charged particle measurement with full and uniform acceptance over the entire azimuth, with pseudorapidity coverage $-0.7\leq \eta\leq 0.7$, so that the jet structure of high $p_T$ scattering could be easily seen and measured. In  Fig.~\ref{fig:mjt-ccorazi}a,b, the azimuthal distributions of associated charged particles 
relative to a $\pi^0$ trigger with transverse momentum $p_{Tt} > 7$ GeV/c are shown for five intervals of associated particle transverse momentum $p_T$. In all cases, strong correlation peaks on flat backgrounds are clearly visible, indicating the di-jet structure which is contained in an interval $\Delta\phi=\pm 60^\circ$ about a direction towards and opposite the to trigger for all values of associated $p_T\, (>0.3$ GeV/c) shown. The width of the peaks about the trigger direction (Fig.~\ref{fig:mjt-ccorazi}a), or opposite to the trigger (Fig.~\ref{fig:mjt-ccorazi}b) indicates out-of-plane activity from the individual fragments of jets. The trigger bias was directly measured from these data by reconstructing the trigger jet from the associated charged particles with $p_T\geq 0.3$ Gev/c, within $\Delta\phi=\pm 60^\circ$ from the trigger particle, using the algorithm $p_{T{\rm jet}}=p_{Tt}+1.5\sum p_T\cos(\Delta\phi)$, where the factor 1.5 corrects the measured charged particles for missing neutrals. The measurements of $\langle z_{\rm trig}\rangle=\langle p_{Tt}/p_{T{\rm jet}}\rangle$ as a function of $p_{Tt}$ for 3 values of $\sqrt{s}$ (Fig.~\ref{fig:mjt-ccormeanz}) show the property of $x_T$ scaling, which was not expected~\cite{JacobEPS79}.  

\begin{figure}[ht]
\begin{center}
\begin{tabular}{cc}
\includegraphics[width=0.47\linewidth]{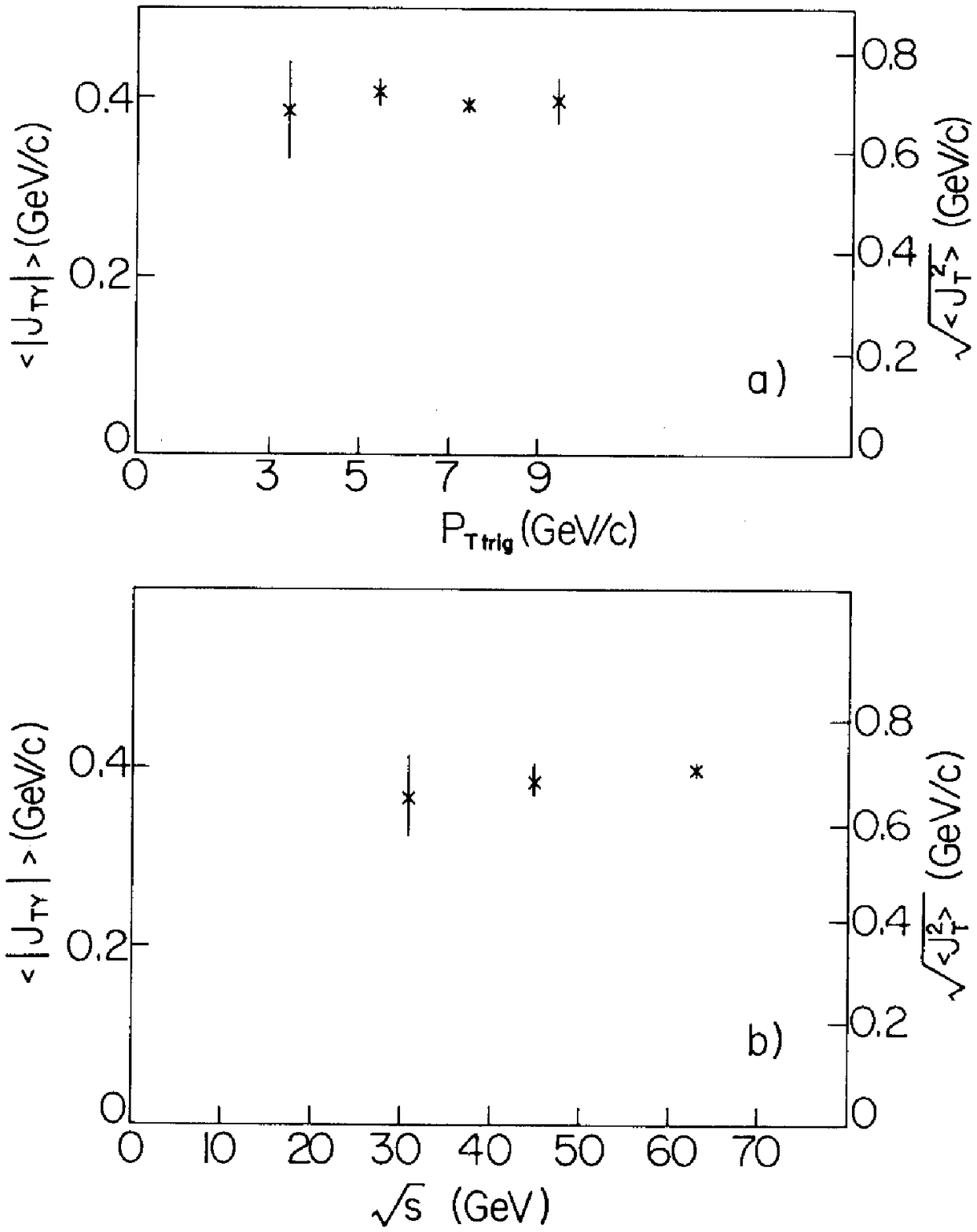} &
\includegraphics[width=0.50\linewidth]{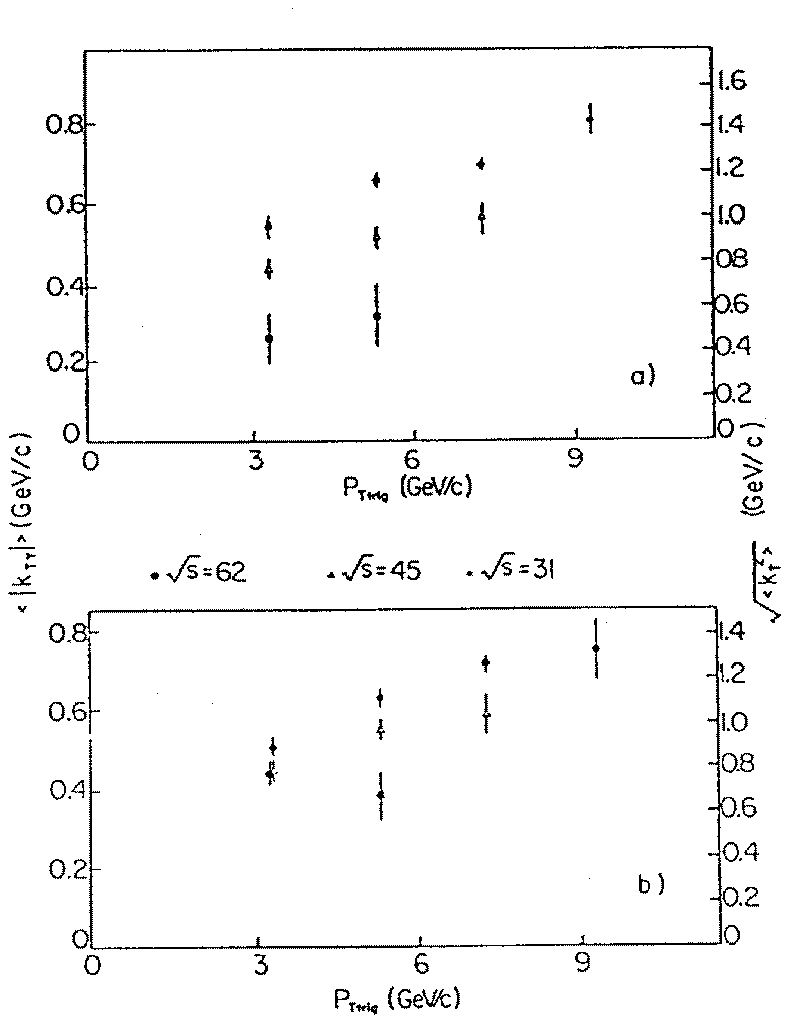}
\end{tabular}
\end{center}
\vspace*{-0.24in}
\caption[]
{CCOR~\cite{CCOR80} measurements of $\langle j_{T_y}\rangle$ (left) $\langle k_{T_y}\rangle$ (right) as a function of $p_{Tt}$ for 3 values of $\sqrt{s}$.   
The mean absolute values of the components $j_{T_y}\equiv j_{T_{\phi}}$ are related to $\sqrt{\langle j_{T}^2}\rangle$ with the assumption that $j_{T_x}$ and $j_{T_y}$ are independent gaussians, with equal r.m.s., which combine independently to form $j_T^2=j_{T_x}^2 + j_{T_y}^2$~\cite{Ap99}.    
\label{fig:mjt-ccorjtkt} }
\end{figure}

  	Following the analysis of previous CERN-ISR experiments~\cite{Darriulat76,CCHK}, the away jet azimuthal angular distributions  of Fig.~\ref{fig:mjt-ccorazi}b, which should be unbiased, were analyzed in terms of the two variables: $p_{\rm out}=p_T \sin(\Delta\phi)$, the out-of-plane transverse momentum of a track;  
 and $x_E$, where:\\ 
\begin{minipage}[c]{0.5\linewidth}
\vspace*{-0.30in}
\begin{equation}	
x_E=\frac{-\vec{p}_T\cdot \vec{p}_{Tt}}{|p_{Tt}|^2}=\frac{-p_T \cos(\Delta\phi)}{p_{Tt}}\simeq \frac {z}{z_{\rm trig}}  
\label{eq:mjt-xE}
\end{equation}
\vspace*{0.06in}
\end{minipage}
\hspace*{0.01\linewidth}
\begin{minipage}[b]{0.50\linewidth}
\vspace*{0.06in}
\includegraphics[scale=0.6]{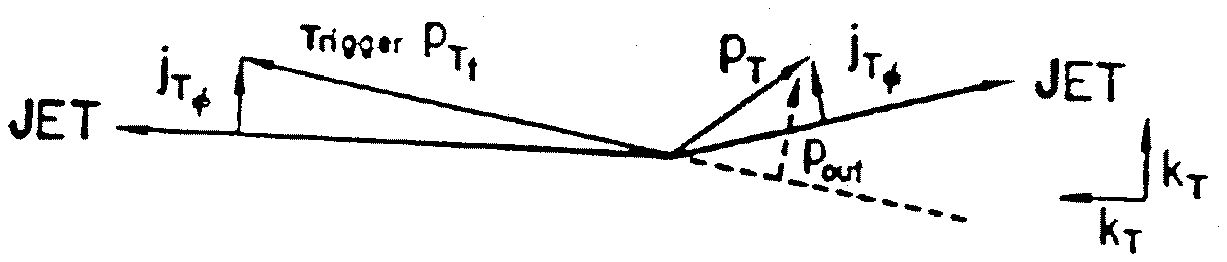}
\vspace*{-0.12in}
\label{fig:mjt-poutxe}
\end{minipage}\vspace*{-0.12in}
$z_{\rm trig}\simeq p_{Tt}/p_{T{\rm jet}}$ is the fragmentation variable of the trigger jet, and $z$ is the fragmentation variable of the away jet. Note that $x_E$ would equal the fragmenation fraction $z$ of the away jet, for $z_{\rm trig}\rightarrow 1$, if the trigger and away jets balanced transverse momentum. 
The $x_E$ distributions~\cite{Angelis79,JacobEPS79b} for the data of Fig.~\ref{fig:mjt-ccorazi}b are shown in Fig.~\ref{fig:mjt-ccorazi}c and show the expected fragmentation behavior, $e^{-6z}\sim e^{-6 x_E \langle z_{\rm trig}\rangle}$. If the width of the away distributions (Fig.~\ref{fig:mjt-ccorazi}b) corresponding to the out of plane activity were due entirely to jet fragmentation, then  
$\langle |\sin(\Delta\phi)|\rangle=\langle |j_{T_{\phi}}|/p_T \rangle$ would decrease in direct proportion to $1/p_T$, where $j_{T_{\phi}}$ is the component of $\vec{j}_T$ in the azimuthal plane, since the jet fragmentation transverse momentum, $\vec{j}_T$, should be independent of $p_T$.  This is clearly not the case, as originally shown by the CCHK collaboration~\cite{CCHK}, which inspired Feynman, Field and Fox (FFF)~\cite{FFF} to introduce, $\vec{k}_T$, the transverse momentum of a parton in a nucleon. In this formulation, the net transverse momentum of an outgoing parton pair is $\sqrt{2} k_T$, which is composed of two orthogonal components, $\sqrt{2} k_{T_{\phi}}$, out of the scattering plane, which makes the jets acoplanar, i.e. not back-to-back in azimuth, and $\sqrt{2} k_{T_x}$, along the axis of the trigger jet, which makes the jets unequal in energy. Originally, ${k}_T$ was thought of as having an `intrinsic' part from confinement, which would be constant as a function of $x$ and $Q^2$, and a part from NLO hard-gluon emission, which would vary with $x$ and $Q^2$, however now it is explained as `resummation' to all orders of QCD~\cite{Sterman}. 
	FFF~\cite{FFF,Levin} gave the approximate formula to derive $k_T$ from the measurement of $p_{\rm out}$ as a function of $x_E$:
\begin{equation}
\langle |p_{\rm out}|\rangle^2=x_E^2 [2\langle |k_{T_{\phi}}|\rangle^2 +  \langle |j_{T_{\phi}}|\rangle^2 ] + \langle |j_{T_{\phi}}|\rangle^2 \qquad .
\label{eq:mjt-FFFpoutkT}
\end{equation}
CCOR~\cite{CCOR80} used this formula to derive $\langle |k_{T_{\phi}}|\rangle$ and $\langle |j_{T_{\phi}}|\rangle$ as a function of $p_{Tt}$ and $\sqrt{s}$ from the data of Fig.~\ref{fig:mjt-ccorazi}b 
(see Fig.~\ref{fig:mjt-ccorjtkt}). This important result shows that $\langle |j_{T_{\phi}}|\rangle$ is constant, independent of $p_{Tt}$ and $\sqrt{s}$, as expected for fragmentation, but that $\langle |k_{T_{\phi}}|\rangle$ varies with both $p_{Tt}$ and $\sqrt{s}$, suggestive of a radiative, rather than an intrinsic origin for $k_T$. 

 \section{Final comments} 
   It should be noted that inclusion of $k_T$ was the key element~\cite{Owens78}  beyond QCD to explain the $n\simeq8$ $x_T$-scaling result of the original CCR~\cite{CCR} high $p_T$ discovery and the FNAL (fixed target) experiments~\cite{Cronin}. More recent FNAL fixed target measurements~\cite{Ap99} and many theoretical works have used $k_T$ as an empirical parameter to improve the comparison of measurement to NLO QCD. However, it is important to remember, as illustrated above, that $k_T$ is not simply a parameter, it can be measured. It is also worthwhile to emphasize that the $k_T$-effect is qualitatively different from NLO QCD. The gaussian nature of $k_T$ which is distinctly different from the NLO power law tail is crisply illustrated {Fig.~\ref{fig:mjt-dmpT} by measurements of the net-transverse momentum distribution of ``Drell-Yan" di-muons produced in p-p collisions~\cite{Ito}, a process which has zero net $p_T$ in LO and diverges in NLO. Another ISR observation~\cite{CCOR82NPB}, not much emphasized there but relevant at  RHIC, is that the measured $\langle z_{\rm trig}\rangle$ is different for single particle inclusive triggers and pair triggers (compare Fig.~\ref{fig:mjt-ccormeanz}-left to Fig.~\ref{fig:mjt-CCORmzpair}). 


\section*{References}
\begin{thebibliography}{99}
\bibitem{BaierQCD98} Baier R 1999 {\em QCD, Proc. IV Workshop-1998 (Paris)}  
(World Scientific, Singapore) pp 272--279   
\bibitem{MJTQCD98}Tannenbaum M J 1999 {\it ibid.}, 
pp 280--285, pp 312--319
\bibitem{Strasbourg} e.g. see 1991 {\em Proc. Int'l Wks. Quark Gluon Plsama 
Signatures (Strasbourg)} (Editions Frontieres, Gif-sur-Yvette, 
France).   
\bibitem{Shura2000} Bazilevsky A {\it et al} 2000 {\it Riken Review No.}\ {\bf 28} 15 
\bibitem{MGyulassy} Gyulassy M and Pl\"umer M  \Journal{\PLB}{243}{432}{1990} 
\bibitem{XNWang} Wang X-N and Gyulassy M \Journal{\PRL}{68}{1480}{1992} 
\bibitem{DIS} Breidenbach M {\it et al}  \Journal{\PRL}{23}{935}{1969}. See also Panofsky W K H 1968 {\em Proc. 14th Int. Conf. HEP (Vienna)} (CERN Scientific Information Service, Geneva, SZ), p. 23
\bibitem{CCR} B\"usser F W {\it et al}  \Journal{\PLB}{46}{471}{1973}, 
see also 1972 {\it Proc. 16th Int. Conf. HEP (Chicago)} (NAL, Batavia, IL) Vol.~3, p.~317 
\bibitem{SS} Banner M {\it et al}  \Journal{\PLB}{44}{537}{1973} 
\bibitem{BS} Alper B {\it et al}  \Journal{\PLB}{44}{521}{1973} 
\bibitem{BBK} Berman S M, Bjorken J D and Kogut J B  
\Journal{\PRD}{4}{3388}{1971}
\bibitem{Bj} Bjorken J D \Journal{\PRD}{179}{1547}{1969}  
\bibitem{CIM} Blankenbecler R,  Brodsky S J and Gunion J F  
\Journal{\PLB}{42}{461}{1972} 
\bibitem{CGKS} Cahalan R F, Geer K A, Kogut J and Susskind L 
\Journal{\PRD}{11}{1199}{1975}
\begin{figure}[!t]
\begin{minipage}{18pc}
\includegraphics[width=18pc,angle=-1]{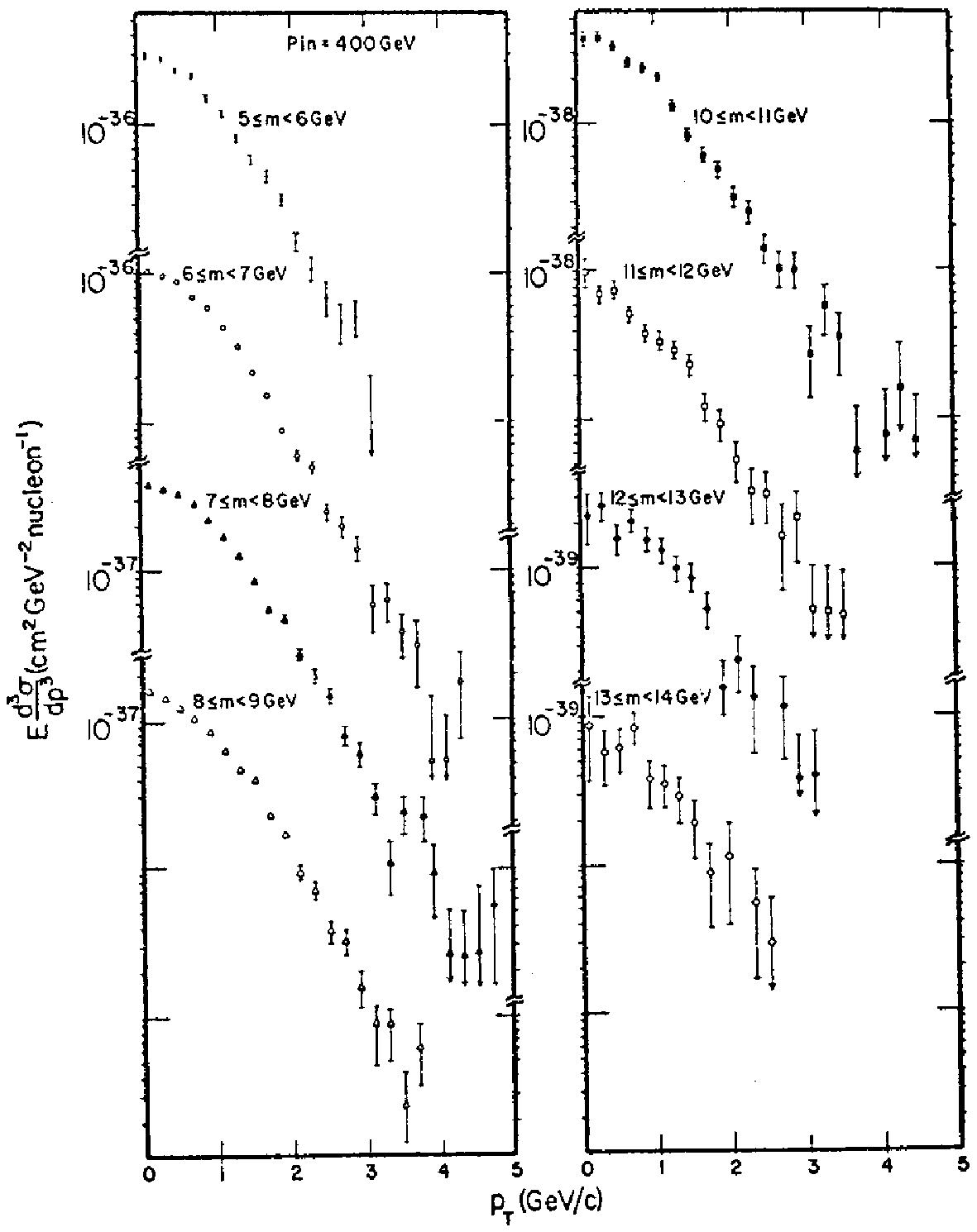}
\caption[]{Invariant yield of di-muons as a function of transverse momentum of the muon pair for 400 GeV incident protons~\cite{Ito}. }
\label{fig:mjt-dmpT}
\end{minipage}\hspace{2pc}%
\begin{minipage}{18pc}
\includegraphics[width=18pc]{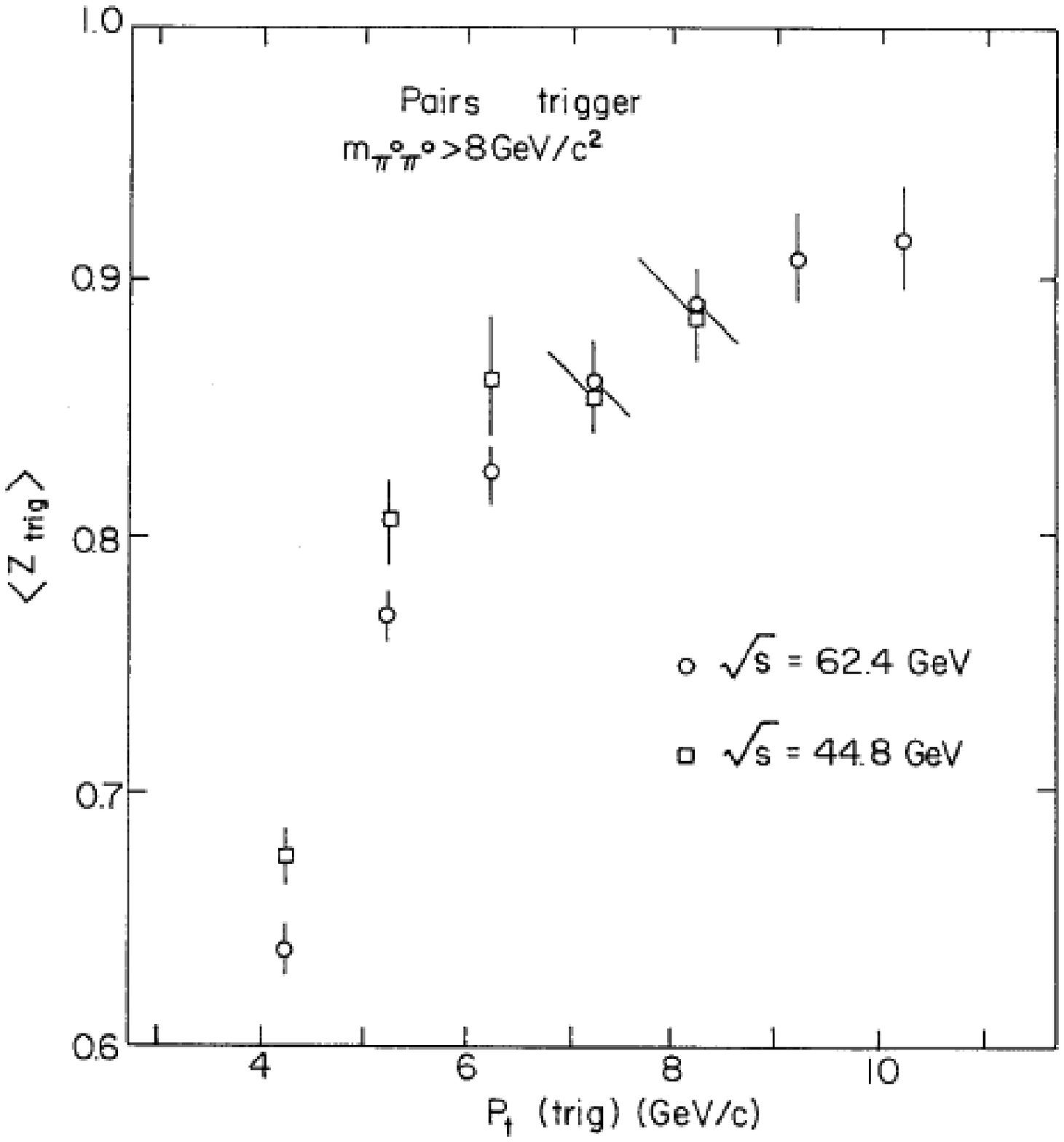}
\caption[]{$\langle z_{\rm trig}\rangle$ for a $\pi^0$-pair trigger with $m_{\pi^0 \pi^0} > 8$ GeV/c$^2$ as a function of $p_{T_t}$~\cite{CCOR82NPB}.  }
\label{fig:mjt-CCORmzpair}
\end{minipage} 
\end{figure}
\bibitem{CCRS} B\"usser F W {\it et al}  
\Journal{\NPB}{106}{1}{1976}
\bibitem{CCOR} Angelis A L S {\it et al}  
\Journal{\PLB}{79}{505}{1978} 
See also, Clark A G {\it et al}  
\Journal{\PLB}{74}{267}{1978} 
\bibitem{Cronin} Antreasyan D, Cronin J W {\it et al}  
\Journal{\PRL}{38}{112}{1977} 
\bibitem{ABCS} Kourkoumelis C {\it et al}  
\Journal{\PLB}{84}{271}{1979}  
\bibitem{CCHK} Della Negra M {\it et al}  
\Journal{\NPB}{127}{1}{1977} 
\bibitem{MJT79} For a contemporary view of the excitement of this period, and some more details, see Tannenbaum M J 1980 {\em Particles and Fields-1979 (Montreal), AIP Conference Proceedings Number 59} (American Institute of Physics, New York) pp. 263-309
\bibitem{FFF} Feynman R P, Field R D and Fox G C \Journal{\NPB}{128}{1}{1977}
\bibitem{Owens78} Owens J F, Reya E and Gl\"uck M 
\Journal{\PRD}{18}{1501}{1978}; Owens J F and Kimel J D  
\Journal{\PRD}{18}{3313}{1978}  
\bibitem{CutlerSivers} Cutler R and Sivers D \Journal{\PRD}{17}{196}{1978}; \Journal{\PRD}{16}{679}{1977}
\bibitem{Combridge:1977dm} Combridge B L, Kripfganz J and Ranft J \Journal{\PLB}{70}{234}{1977}
\bibitem{Gordon} {\AA}kesson T {\it et al}  
\Journal{\PLB}{128}{354}{1983}  
\bibitem{MJTIJMPA} e.g. for a review, see Tannenbaum M J  
\Journal{\IJMPA}{4}{3377}{1989} 
\bibitem{Paris82} 1982 {\it J. Phys.}  C {\bf 3} {\em Proc. 21st Int'l Conf. HEP (Paris)}: see Repellin J P p.  
C3-571; also see Tannenbaum M J p. C3-134, Wolf G p. C3-525  
\bibitem{CCOR82NPB} Angelis A L S {\it et al}  
\Journal{\NPB}{209}{284}{1982}
\bibitem{Owens} Owens J F \Journal{\RMP}{59}{465}{1987} 
\bibitem{Darriulat} Darriulat P \Journal{\ARNS}{30}{159}{1980}
\bibitem{DiLella} DiLella L \Journal{\ARNS}{35}{107}{1985}
\bibitem{confusion} However, this delicacy is often glibly ignored 
\bibitem{JacobLandshoff} Jacob M and Landshoff P \Journal{\PLC}{48}{286}{1978} 
\bibitem{Moriond79} e.g. see 1979 {\em Proc. XIV Rencontre de Moriond---``Quarks, Gluons and Jets" (Les Arcs)} (Editions Fronti\`eres, Dreux, France)  Boggild H  p. 321, Tannenbaum M J p351, and references therein  
\bibitem{Angelis79} Angelis A L S {\it et al} 1979 {\it Physica Scripta} {\bf 19}\ 116
\bibitem{JacobEPS79} Jacob M 1979 {\em Proc. EPS Int'l Conf. HEP (Geneva)} (CERN, Geneva) Volume 2, pp. 473-522 
\bibitem{Darriulat76} Darriulat P {\it et al}  \Journal{\NPB}{107}{429}{1976} 
\bibitem{JacobEPS79b} See p. 512 in reference \cite{JacobEPS79} 
\bibitem{Sterman} e.g. see Kulesza A, Sterman G and Vogelsang W \Journal{\PRD}{66}{014011}{2002}
\bibitem{Levin} See also, Levin E M and Ryskin M G 1975 {\it Sov. Phys. JETP}  {\bf 42} \ 783
\bibitem{CCOR80} Angelis A L S {\it et al}  \Journal{\PLB}{97}{163}{1980} 
\bibitem{Ap99} Apanasevich L {\it et al}  \Journal{\PRD}{59}{074007}{1999}
\bibitem{Ito} Ito A S {\it et al}  \Journal{\PRD}{23}{604}{1981} 

\end{thebibliography}
\end{document}